\documentclass[twocolumn,showpacs,preprintnumbers,amsmath,amssymb]{revtex4}

\usepackage{graphicx}
\usepackage{dcolumn}
\usepackage{bm}
\usepackage{amsmath}

\begin{document}
\title{Traffic flow and efficient routing on scale-free networks: A survey}
\author{Bing-Hong Wang}
\author{Tao Zhou}
\email{zhutou@ustc.edu}

\affiliation{%
Department of Modern Physics and Nonlinear Science Center,
University of Science and Technology of China, Hefei 230026, PR
China
}%

\date{\today}

\begin{abstract}
Recently, motivated by the pioneer works in revealing the
small-world effect and scale-free property of various real-life
networks, many scientists devote themselves to studying complex
networks. In this paper, we give a brief review on the studies of
traffic flow and efficient routing on scale-free networks,
including the traffic dynamics based on global routing protocol,
Traffic dynamics based on local routing protocol, and the critical
phenomena and scaling behaviors of real and artificial traffic.
Finally, perspectives and some interesting problems are proposed.
\end{abstract}
\pacs{89.75.Hc,89.20.Hh,05.10.-a,89.40.-a}

\maketitle

\section{Introduction}
Many social, biological, and communication systems can be properly
described as complex networks with vertices representing
individuals or organizations and links mimicking the interactions
among them
\cite{Albert2002,Dorogovtsev2002,Newman2003,Boccaletti2006}. One
of the ultimate goals of the current studies on topological
structures of networks is to understand and explain the workings
of systems built upon those networks, for instance, to understand
how the topology of the World Wide Web affects Web surfing and
search engines, how the structure of social networks affects the
spread of diseases, information, rumors or other things, how the
structure of a food web affects population dynamics, and so on.
The increasing importance of large communication networks such as
the Internet \cite{Pastor2004}, upon which our society survives,
calls for the need for high efficiency in handling and delivering
information. Therefore, to understand the traffic dynamics and
find the optimal strategies for traffic routing is one of the
important issues we have to address. There have been many previous
studies to understand and control traffic congestion on networks,
with a basic assumption that the network has a homogeneous
structure
\cite{Li1989,Leland1993,Taqqu1997,Crovella1997,Arenas2001}.
However, many real-life communication networks, such as the
Internet \cite{Pastor2001} and the World-Wide-Web
\cite{Albert1999}, display scale-free degree distribution
\cite{Barabasi1999a,Barabasi1999b}, thus it is of great interest
to study the traffic flow on scale-free networks. In this light,
the traffic dynamics on complex networks have recently attracted a
large amount of interest from the physical community.

In this paper, we will give a brief review about traffic dynamics
on scale-free networks. This paper is organized as follow: In Sec.
2 and 3, the traffic dynamics with global and local routing
protocols are introduced, respectively. In Sec. 4, the critical
phenomena and self-similarity scaling of real traffic and
artificial models are discussed. Finally, we sum up this paper in
Sec. 5.

\section{Traffic dynamics based on global routing protocol}
In this section, we discuss the case where the whole topological
information is available for each router. For simplicity, all the
nodes are treated as both hosts and routers. The simplest model
can be described as follows:

(1) At each time step, there are $R$ packets generated in the
system, with randomly chosen sources and destinations. Once a
packet is created, it is placed at the end of the queue if this
node already has several packets waiting to be delivered to their
destinations.

(2) At each time step, each node, $i$, can deliver at most $C_i$
packets one step toward their destinations according to the
routing strategy.

(3) A packet, once reaching its destination, is removed from the
system.

We are most interested in the critical value $R_c$ where a phase
transition takes place from free flow to congested traffic. This
critical value can best reflect the maximum capability of a system
handling its traffic. In particular, for $R<R_c$, the numbers of
created and delivered packets are balanced, leading to a steady
free traffic flow. For $R>R_c$, traffic congestion occurs as the
number of accumulated packets increases with time, simply for that
the capacities of nodes for delivering packets are limited. To
characterize the phase transition, we use the following order
parameter
\begin{equation}
H(R)=\lim_{t \rightarrow \infty} \frac{C}{R} \frac{\langle \Delta
W \rangle}{\Delta t},
\end{equation}
where $\Delta W = W(t+\Delta t) - W(t)$, with $\langle \cdots
\rangle$ indicating average over time windows of width $\Delta t$,
and $W(t)$ is the total number of packets in the network at time
$t$. Clearly, $H$ equals zero in the free-flow state, and becomes
positive when $R$ exceeds $R_c$.

\begin{figure}
\scalebox{0.8}[0.8]{\includegraphics{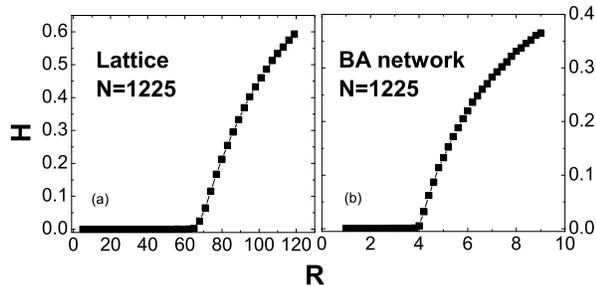}}
\caption{\label{fig:epsart} The order parameter $H$ versus $R$ for
two-dimensional lattice (a) and BA networks (b) with the same size
$N=1225$ and average degree $\langle k\rangle=4$. The delivering
capability of each node is simply set as a constant $C=1$. The
shortest-path routing strategy yields
$R_c^{\texttt{Lattice}}\approx60$ and
$R_c^{\texttt{BA}}\approx4.0$, respectively.}
\end{figure}

Since in the Internet, the deviation of traffic from the shortest
path is only about $10\%$ \cite{Krioukov2003}, one can assume that
the routing process takes place according to the criterion of the
shortest available path from a given source to its destination.
Accordingly, firstly, we investigate the shortest-path routing
strategy, which can be implemented by either of the two ways,
finding the shortest-path dynamically or following the fixed
routing table. In the former case \cite{Zhao2005}, for each newly
generated packet, the router will find a shortest path between its
source and destination, and then, the packet is forwarded along
this path during the following time steps. In the latter case
\cite{Zhou2006a}, for any pair of source and destination, one of
all the shortest paths between them is randomly chosen and put
into the fixed routing table that is followed by all the
information packets. Compared with the dynamical routing algorithm
and the information feed-back mechanism, the fixed routing
algorithm is much more widely used in real communication systems
for its obvious advantages in economical and technical costs
\cite{Tanenbaum1996,Huitema2000}. Actually, the behaviors of those
two cases are pretty much the same \cite{Zhao2005,Zhou2006a}, thus
we will not distinguish them hereinafter.

If the delivering capability of each node is the same, the
critical point $R_c$ of highly heterogeneous network will be much
smaller than that of homogeneous network. It is because that when
all the packets follow their shortest paths, it will easily lead
to the overload of the heavily-linked router, and the congestion
will immediately spread over all the nodes. Fig. 1 shows the order
parameter $H$ versus $R$ for (a) the two-dimensional lattice with
periodical boundary condition and (b) the Barab\'asi-Albert (BA)
network \cite{Barabasi1999a,Barabasi1999b} with average degree
$\langle k \rangle=4$. Clearly, the throughput, measured by the
$R_c$, of regular network is much larger than that of scale-free
networks.

To provide the theoretical estimate of the value of $R_c$, we
first introduce the concept of betweenness centrality (see also
the original concept of centrality
\cite{Anthonisse1971,Freeman1977}, the generalized concept of
centrality \cite{Scott2000}, the physical meaning of betweenness
centrality \cite{Newman2001}, and some recently proposed
algorithms for calculating betweenness
\cite{Brandes2001,Newman2004,Zhou2006b}). The betweenness
centrality of a node $v$ is defined as
\begin{equation}
B_v=\sum_{s\neq t}\frac{\sigma_{st}(v)}{\sigma_{st}},
\end{equation}
where $\sigma_{st}$ is the number of shortest paths going from $s$
to $t$ and $\sigma_{st}(v)$ is the number of shortest paths going
from $s$ to $t$ and passing through $v$. Below the critical value
$R_c$, there is no accumulation at any node in the network and the
number of packets that arrive at node $v$ is, on average,
$RB_v/N(N-1)$. Therefore, a particular node will collapse when
$RB_v/N(N-1)>C_v$. Considering the transferring capacity of each
node is fixed to $C$ and congestion occurs at the node with the
largest betweenness centrality, $R_c$ can be estimated as
\cite{Zhao2005,Guimera2002}
\begin{equation}
R_c=CN(N-1)/B_{\texttt{max}},
\end{equation}
where $B_{\texttt{max}}$ is the largest betweenness centrality of
the network. This equation illuminates that the network of larger
$B_{\texttt{max}}$ has smaller throughput.

For many real-life networks, the betweenness centrality is
strongly correlated with degree. In general, the larger the
degree, the larger the centrality. For many scale-free networks,
it has been shown that the betweenness centrality approximately
scales as $B(k)\sim k^\mu$ \cite{Goh2001,Barthelemy2003}, where
$B(k)$ denotes the average betweenness centrality over all the
$k$-degree nodes. Therefore, in a heterogeneous network, there
exists a few high-betweenness nodes, named \emph{hub nodes}, which
are easy to be congested. This is precisely the cause of low
throughput of scale-free networks.

To enhance the traffic capability, Zhao \emph{et al.} proposed two
traffic models \cite{Zhao2005}, where the delivering capability of
a node $i$ is assigned as $C_i=1+\beta k_i$ (Model I) and
$C_i=1+\beta B_i$ (Model II), respectively. Here, $0<\beta<1$ is a
control parameter. As we have mentioned above (see Eq. (3)), it is
clear that the throughput of the whole network will increase if
the hub nodes have higher delivering capability though the total
capability $\sum_i C_i$ keeps unchanged. This work suggests a way
to alleviate traffic congestions for highly heterogeneous
networks: making nodes with large betweenness as powerful and
efficient as possible for processing and transmitting information.
Particularly, in the model II, the throughput $R_c$ is independent
of the network topology.

\begin{figure}
\scalebox{0.40}[0.50]{\includegraphics{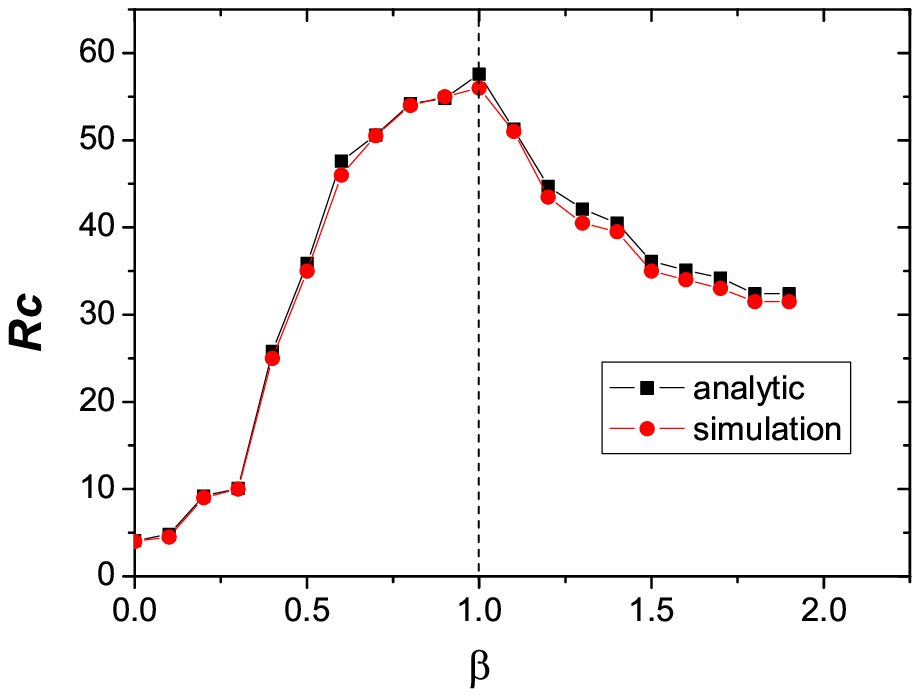}}
\scalebox{0.40}[0.50]{\includegraphics{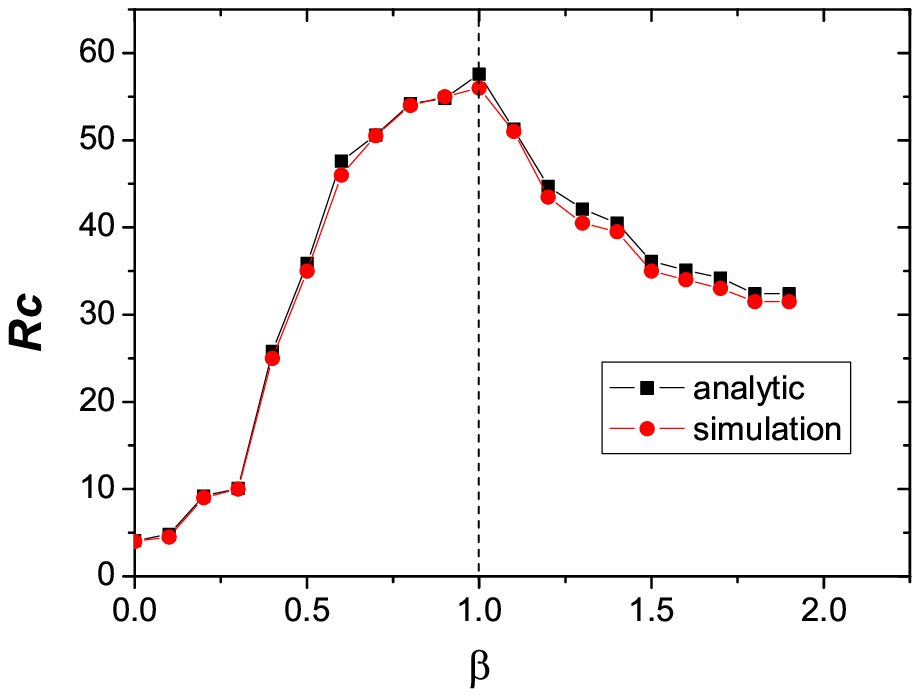}}
\caption{\label{fig:epsart} (Color online) The critical $R_{c}$ vs
$\beta$ for BA network with average degree $\langle k \rangle=4$
and of size $N=1225$ (left panel) and $N=1500$ (right panel). Both
simulation and analysis demonstrate that the maximal $R_{c}$
corresponds to $\beta \approx 1$. The results are the average over
10 independent runs.}
\end{figure}

However, there are two reasons that hinder the application of
those models. First, the capability/power distributions in some
real networks are homogeneous, although their degree distributions
are heterogeneous \cite{Zhou2006c}. For example, in the
broadcasting networks, the forwarding capacity of each node is
limited. Especially, in wireless multihop ad hoc networks, each
node usually has the same power thus almost the same forwarding
capacity \cite{Gupta2000}. Second, the structure of real networks
evolve ceaselessly, thus the degree and betweenness centrality of
each node vary ever and again. By contrary, one can not change the
delivering capability of a node freely due to the technical
limitation.

For the case that each node has the same delivering capability
$C$, Yan \emph{et al.} proposed a novel routing algorithm to
enhance the network throughput \cite{Yan2006,Zhou2006d}. Note
that, the path with shortest length is not necessarily the
quickest way, considering the presence of possible traffic
congestion and waiting time along the shortest path. Obviously,
nodes with larger degree are more likely to bear traffic
congestion, thus a packet will by average spends more waiting time
to pass through a high-degree node. All too often, bypassing those
high-degree nodes, a packet may reach its destination quicker than
taking the shortest path. For any path between nodes $i$ and $j$
as $P(i\rightarrow j):=i\equiv x_0, x_1, \cdots x_{n-1}, x_n
\equiv j$, denote
\begin{equation}
L(P(i\rightarrow j):\beta)=\sum_{i=0}^{n-1}k(x_i)^{\beta},
\end{equation}
where $\beta$ is a tunable parameter. The \emph{efficient path}
between $i$ and $j$ is corresponding to the route that makes the
sum $L(P(i\rightarrow j):\beta)$ minimum. Obviously,
$L_{min}(\beta=0)$ recovers the traditionally shortest path
length. All the information packets follow the efficient paths
instead of the shortest paths.

In Fig. 2, we report the simulation results for the critical value
$R_{c}$ as a function of $\beta$ on BA network with the size
$N=1225$ and $N=1500$, which demonstrate that the optimal router
strategy corresponding to $\beta=1$ and the size of BA network
doesn't affect the value of optimal $\beta$. In comparison with
the traditional routing strategy (i.e. $\beta=0$), the throughput
$R_c$ of the whole network is greatly improved more then 10 times
without any increase in algorithmic complexity. By extending the
concept of betweenness centrality to efficient betweenness
centrality, that is, using the efficient paths instead of the
shortest paths in the definition of betweenness centrality, the
analytic results can be obtained according to the Little's law
\cite{Zhao2005,Guimera2002,Yan2006}. The analytical results are
also shown in Fig. 2, which agree very well with the simulations.
In the previous studies, the betweenness centrality is always
considered as a static topological measure of networks under the
latent assumption that all the information packets go along the
shortest paths from source to destination. The work of Yan
\emph{et al.} shows that this quantity (efficient betweenness) is
determined both by the routing algorithm and network topology,
thus one should pay more attention to the design of routing
strategies. For example, a more intelligent router that can detour
at obstacle performs much better than traditional router which
just waits at obstacle \cite{Holme2003}, and a recent work
demonstrates that the dynamical information can be used to design
more efficient routing strategy \cite{Chen2006}.

\section{Traffic dynamics based on local routing protocol}
Although the routing protocol using global topological information
is very efficient, it is not practical for huge-size communication
networks and the evolving networks since the technical limitation
of the router. It is because the router hardware is hard to be
designed to have the capability to storage much information or
adapt dynamical information. Therefore, it is also very
interesting and practical to study the traffic behaviors on
scale-free networks based on local information. The simplest
network traffic model on local protocol is the random-walk
process, where each packet is delivered to randomly selected one
of its neighbors as far as it reaches the destination. The
random-walk on scale-free networks has been extensively explored
previously \cite{Noh2004,Eisler2005,Noh2006}, however, it is far
from the real traffic since it can not reproduce the self-similar
scaling as we will present in the next section.

\begin{figure}
\scalebox{0.7}[0.7]{\includegraphics{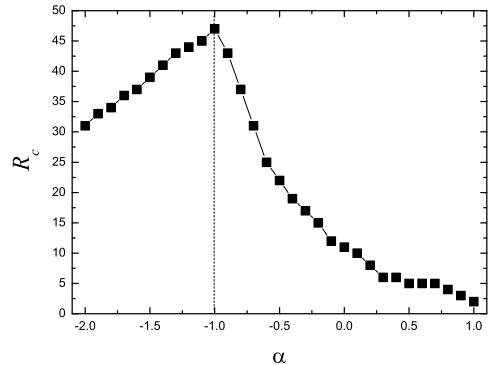}}
\caption{\label{fig:epsart} The critical $R_{c}$ vs $\alpha$ for
BA network with average degree $\langle k \rangle=10$ and of size
$N=1000$. The delivering capability of each node is set as a
constant $C=10$. The results are the average over 10 independent
runs.}
\end{figure}

Motivated by the previous studies about searching engine
\cite{Adamic2001,Kim2002} and the global routing strategy
\cite{Yan2006} on complex networks, Yin \emph{et al.} proposed a
traffic model using preferential selection among local neighbors
\cite{Yin2006}. In this model, to navigate packets, each node
performs a local search among its neighbors. If a packet's
destination is neighboring, it will be delivered directly to its
target, otherwise, it will be forwarded to a neighbor $j$ of its
currently located node $i$ according to the preferential
probability
\begin{equation}
\Pi_{i\rightarrow j}=\frac{k_j^\alpha}{\sum_l k_l^\alpha},
\end{equation}
where the sum runs over the neighbors of node $i$, $k_i$ is the
degree of node $i$, and $\alpha$ is an adjustable parameter.
Similar to the models mentioned in the last section, the
first-in-first-out (FIFO) discipline is applied at the queue of
each node. Another important rule, named path iteration avoidance
(PIA), is that any edge cannot be visited more than twice by the
same packet. Set the delivering capability of each node as a
constant $C$, the simulation results show that the optimal
performance of the whole system corresponds to $\alpha=1$ (see
Fig. 3). This optimal value can also be analytically obtained
\cite{Wang2006}. Further more, if the delivering capability of
each node is proportional to its degree, the optimal value of
$\alpha$ will shift to $\alpha=0$ \cite{Wang2006}.

\begin{figure}
\begin{center}
\scalebox{1}[0.85]{\includegraphics{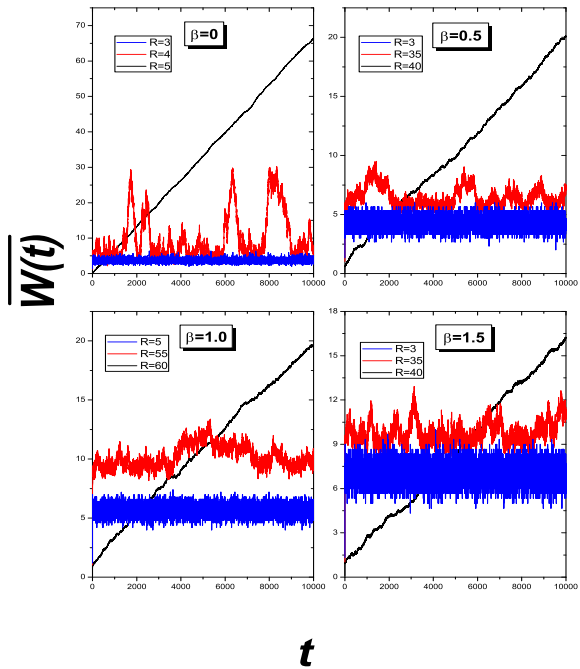}} \caption{(Color
online) The traffic rate process for free (red), critical (blue)
and congested (black) states with different $\beta$. All the data
are obtained from an artificial traffic model \cite{Yan2006}.}
\end{center}
\end{figure}

It is worthwhile to emphasize that the behavior of Yin \emph{et
al.}'s model \cite{Yin2006} is similar to that of Yan \emph{et
al.}'s model \cite{Yan2006}, and the behavior of Wang \emph{et
al.}'s model \cite{Wang2006} is similar to that of Zhao \emph{et
al.}'s model \cite{Zhao2005}. These resemblances indicate the
existence of some common policies between the design of global and
local routing protocols, that is, to bypass the hub nodes or to
improve the delivering capability of these nodes can sharply
enhance the throughput of the whole network.

Note that, each router in the present models
\cite{Yin2006,Wang2006} needs to know all its neighbors' degrees
and a packet has to remember the links its has visited, which
requires much intelligence of the system. It may damage the
advantage of local routing strategy since to implement the PIA
rule will make this system even more complicated than the one
using fixed routing algorithm. And the throughput of networks is
very low without the PIA rule due to many times of unnecessary
visiting along the same links by the same packets.

Another factor that affects the performance of local routing
strategy is the area of information a router can make use of.
Based on an artificial directed World-Wide-Web model (see some
recently proposed theoretical models of directed World-Wide-Web
\cite{Tadic2001,Liu2006}), Tadi\'c \emph{et al.} investigated a
local routing protocol with finite buffer size of each router, and
found that the next-to-nearest routing algorithm can perform much
better than the nearest routing algorithm \cite{Tadic2002}. In
this model each packet follows a random-walk unless its
destination appears within the current router's searching area,
and the next-to-nearest routing algorithm means the router can
directly deliver a packet to its destination if this destination
is within two steps.

\section{The critical phenomena and scaling behaviors of traffic}
Recent empirical studies on communication network have found
pervasive evidence of some surprising scaling properties. one
example of such discoveries is that the traffic rates of both a
given link in the Internet and a local Ethernet exhibit the
self-similarity (or fractal-like) scaling, and the multifractal
scaling is also found over small time scale
\cite{Crovella1997,Leland1994,Paxson1997,Feldmann1999,Yang2006}.
These empirical studies describe pertinent statistical
characteristics of temporal dynamics of measured traffic rate
process and provide ample evidence that these traces are
consistent with long-range correlated behavior. Furthermore, the
observation of a phase transition between the free-flow phase and
the congested phase in the Internet traffic is demonstrated by
Takayasu \emph{et al.} through both the round trip time experiment
\cite{Takayasu1996,Fukuda1999} and packet flow fluctuation
analysis \cite{Takayasu1999,Takayasu2000}. They found that the
real traffic exhibits the $1/f$-type scaling, however, this $1/f$
scaling can only be observed near the critical state
\cite{Takayasu1996,Fukuda1999,Takayasu1999,Takayasu2000}.

Cai \emph{et al.} \cite{Cai2006a} investigated the scaling
behavior of mimic traffic rate process near at the critical point
generated by an the model of Yan \emph{et al.}
\cite{Yan2006,Zhou2006d}. Fig. 4 reports the average number of
packets over all the nodes, $\overline{W}(t)=W(t)/N$, as a time
series in free, critical and congested states, respectively. The
behaviors of $\overline{W}(t)$ in the free and congested states
are very simple: In the former case, it fluctuates slightly around
a very low value, while in the latter case, it increases linearly.
However, the time series at the critical point is very
complicated, which exhibits some large fluctuations like those
have been observed in the real traffic \cite{Field2004}. This
reason resulting in this phenomenon may be the usage of global
routing strategy that leads to a possible long-range correlation,
since this phenomenon can not be detected from the random-walk
model \cite{Noh2004} and the model based on local protocol
\cite{Yin2006,Wang2006}. However, a very similar phenomenon is
also observed in a traffic model with local protocol
\cite{Tadic2004}, where the buffer size of each router is finite.

\begin{figure}
\begin{center}
\scalebox{0.7}[0.7]{\includegraphics{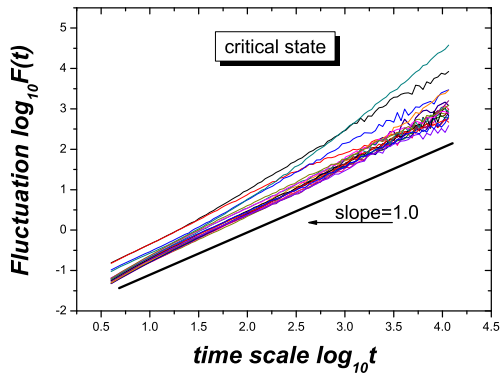}} \caption{(Color
online) The detrended fluctuation analysis of the traffic rate
processes generated by an artificial traffic model \cite{Yan2006}.
All the data are obtained from the critical state, and the
different curves represent the cases of different $\beta$ from 0
to 1.9 at step 0.1.}
\end{center}
\end{figure}

In the previous studies, the autocorrelation function and power
spectrum are widely used to analyse the self-similar scaling
behaviors of both real \cite{Holt2000,Karagiannis2004} and
artificial \cite{Tadic2004} traffic data. However, it is shown
that all the above methods do not work very well for the effect of
non-stationary \cite{Liu1999}, and are less accurate than the
detrended fluctuation analysis (DFA) proposed by Peng \emph{et
al.} \cite{Peng1994,Peng1995}, which has now been accepted as an
important time series analysis approach and widely used especially
for financial and biological data
\cite{Bunde2000,Yang2004,Xu2005,Chen2005,Cai2006b}.

The DFA method is based on the idea that a correlated time series
can be mapped to a self-similar process by an integration.
Therefore, measuring the self-similar feature can indirectly tell
us the information about the correlation properties. Briefly, the
description of the DFA algorithm involves the following steps.

(1) Consider a time series $x_{i}, i=1, \cdots, N$, where $N$ is
the length of this series. Determine the \emph{profile}
\begin{equation}
y(i)=\sum_{k=1}^{i}[x_{k}-\langle x\rangle], i=1, \cdots, N,
\end{equation}
where
\begin{equation}
\langle x\rangle=\frac{1}{N}\sum_{i=1}^{N}x_{i}.
\end{equation}

(2) Divide profile $y(i)$ into non-overlapping boxes of equal size
$t$.

(3) Calculate the local trend $y_{\texttt{fit}}$ in each box of
size $t$ by a least-square fit of the series, and the detrended
fluctuation function is given as
\begin{equation}
Y(k)= y(k)-y_{\texttt{fit}}(k).
\end{equation}

(4) For a given box size $t$, we calculate the root-mean-square
fluctuation
\begin{equation}
F(t)=\sqrt{\frac{1}{N}\sum_{1}^{N}[Y(k)]^{2}},
\end{equation}
and repeat the above computation for different box sizes $t$ (i.e.
different scales) to provide a relationship between $F$ and $t$.
If the curve $F(t)$ in a log-log plot displays a straight line,
then $F(t)$ obeys the power-law form $t^{H}$ with $H$ the slope of
this line.

As shown in Fig. 5, the traffic rate time series generated by the
model of Yan \emph{et al.} \cite{Yan2006} also exhibits the
self-similar scaling behaviors at the critical point of phase
transition. The scaling exponents calculated with DFA for
different $\beta$ are approximate $H\approx 1$, and the value of
$\beta$ has almost no effect on $H$. This value of $H$ implies the
$1/f$-type scaling in the power spectrum and the long-range
correlated behavior in a wide range of scales. A very recent
empirical study on the traffic rate process of a University
Ethernet has demonstrated that the real Ethernet traffic displays
a self-similarity behavior with scaling exponent $\approx 0.98$
\cite{ZhouPL2006}, which agrees well with the present result
$H\approx 1$.

\section{Conclusion and Discussion}
The studies of network traffic are now in the ascendant. Many new
discoveries, especially the role of network heterogeneity that can
sharply reduce the traffic capacity, provide us with new scenarios
and problems in understanding and controlling traffic congestions.
Physicists being not only the new object named \emph{scale-free
networks}, but also the new methodologies much different from
those usually used in the engineering science. As an end of this
brief review, we list a few interesting problems still unsolved
below.

\emph{Problem 1:} The visual field of router may be one of the
most important factors that affects the traffic capacity of whole
networks. In the random walk \cite{Noh2004} the router knows
nothing about the topological information; in the preferential
routing strategy \cite{Yin2006}, the router knows the topological
information of all its nearest neighbors; in the global routing
protocol \cite{Zhao2005,Yan2006}, each router knows the whole
topological information. Clearly, with wider visual field, the
system can perform better. Tadi\'c and Rodgers \cite{Tadic2002}
proposed a local traffic model where each router knows the
topological information of all its nearest and next-nearest
neighbors, which, as expected, has obviously higher throughput
than the case where only the nearest neighbors' information is
available. Up to now, it lacks a systemic study on the effect of
router's visual field on the traffic condition of networks, which
may worth some further work.

\emph{Problem 2:} A router can memorize huge information about
shortest or efficient paths that, at least, can be used to
implement the strategy of fixed routing table is very expensive.
Even worth, the current technology does not support the router to
do dynamical path-finding in large-size networks. So, a relative
problem to the preceding one is that what will happen if one mixes
the global and local protocols together, that is, a few routers in
the networks can do global path-finding or have memorized the
shortest/efiicient paths and others can only perform local
protocol. A further question is that if the addition of a few very
powerful routers to a traffic system based on local protocol can
sharply enhance the network throughput, which locations should
these powerful routers choose?

\emph{Problem 3:} Although $\beta=1$ corresponds to the optimal
value of network throughput when using efficient-path finding
strategy \cite{Yan2006}, it is really a bad strategy when the
traffic density is sparse since to bypass the hub nodes will
increase the average distance between source and destination. If
the traffic density is sparse, the strategy with $\beta>0$ will
waste time. Therefore, a natural question raises: How to use the
dynamical information to improve the performance of network
traffic? Can we design some on-line algorithms to guide the
information packets?

\emph{Problem 4:} On one hand, in the network traffic dynamics,
the maximal betweenness centrality $B_\texttt{max}$ is the key
factor that determines the network throughput $R_c$ since the node
having maximal betweeness centrality is the bottleneck of
information traffic thus is firstly congested. On the other hand,
the node having maximal betweeness centrality is also the
bottleneck that hinders the synchronization signal's
communication, thus the network of higher $B_\texttt{max}$ may of
poorer synchronizability \cite{Nishikawa2004,Hong2004}. Therefore,
we guess there may exist some common features between network
traffic and network synchronization, although they seem completely
irrelevant. Actually, some recently proposed methods used to
enhance the network synchronizability can also be used to enhance
the network throughput
\cite{Motter2005,Chavez2005,Zhao2005b,Zhou2006e}. An in-depth
investigation is of great theoretical interest and we want to know
if those two different kinds of dynamics, traffic and
synchronization, belonging to some kind of ``universality class".

\emph{Problem 5:} The routing strategies for real Internet
\cite{Echenique2004} is of special interest for its practical
significance. However, the real Internet is highly clustered and
displaying hierarchical structure \cite{Pastor2004}, thus far from
the extensively studied BA networks. Although there exists some
highly-clustered models with hierarchical structures
\cite{Holme2002,Zhou2005}, they can not capture the detailed
topological properties of real Internet. We have noticed that a
recent model \cite{Zhou2004} aiming at Internet is very close to
the reality, thus it is interesting to explore the difference
between simulation results based on BA networks and the model of
Zhou and Mondrag\'on \cite{Zhou2004}.

\emph{Problem 6:} The previous studies mainly focus on the
information flow and corresponding routing strategies. However, in
the urban traffic, it is not the routes but the drivers are
intelligent. How can they make use of traffic information to
shorten their travelling time \cite{Wang2005}, and whether the
selfish of each agent will reduce the system profit
\cite{Youn2006}?

\begin{acknowledgments}
BHWang acknowledges the support of the National Basic Research
Program of China (973 Program) under Grant No. 2006CB705500, the
Special Research Founds for Theoretical Physics Frontier Problems
under Grant No. A0524701, the Specialized Program under the
Presidential Funds of the Chinese Academy of Science, and the
National Natural Science Foundation of China under Grant Nos.
10472116, 10532060, and 10547004. TZhou acknowledges the support
of the National Natural Science Foundation of China under Grant
Nos. 70471033, 70571074, and 70571075, and the Graduate Student
Foundation of University of Science and Technology of China under
Grant Nos. KD2004008 and KD2005007.
\end{acknowledgments}

\end{document}